\documentclass[aps,prb,showpacs,twocolumn]{revtex4-1}

\usepackage{amsmath,graphics}
\usepackage[next]{inputenc}
\usepackage[dvips]{epsfig}

\def\be{\begin{equation}}
\def\ee{\end{equation}}

\def\bi{\begin{itemize}}
\def\ei{\end{itemize}}
\def\bn{\begin{enumerate}}
\def\en{\end{enumerate}}
\def\bea{\begin{eqnarray}}
\def\eea{\end{eqnarray}}

\def\ba{\begin{array}}
\def\ea{\end{array}}
\def\bd{\begin{displaymath}}
\def\ed{\end{displaymath}}

\begin{document}
\title{Competing Exotic Topological Insulator Phases in Transition Metal Oxides on the Pyrochlore Lattice with Distortion}
\author{Mehdi Kargarian}
\email[]{kargarian@physics.utexas.edu}
\affiliation{Department of Physics, The University of Texas at Austin, Austin, TX 78712, USA}
\author{Jun Wen}
\email[]{jwen@physics.utexas.edu}
\affiliation{Department of Physics, The University of Texas at Austin, Austin, TX 78712, USA}
\author{Gregory A. Fiete}
\affiliation{Department of Physics, The University of Texas at Austin, Austin, TX 78712, USA}

\begin{abstract}
In this work we investigate the phase diagram of heavy ($4d$ and $5d$) transition metal oxides on the pyrochlore lattice, such as those of the form $\mathrm{A_2M_2O_7}$, where A is a rare earth element and M is a transition metal element. We focus on the competition between Coulomb interaction, spin-orbit coupling, and lattice distortion when these energy scales are comparable. Strong spin-orbit coupling entangles the spin and the $t_{2g}$ $d$-orbitals giving rise to doublet $j=1/2$ and quadruplet $j=3/2$ states.  In contrast to  previous works which focused on the doublet manifold, we also discuss the quadruplet manifold which is relevant for several pyrochlore oxides. The Coulomb interaction is taken into account by use of the slave-rotor mean field theory and different classes of lattice distortions which further split the levels of the quadruplet $j=3/2$ manifold are studied. Various topological phases are predicted, including exotic strong and weak topological Mott insulating phases.  We discuss the general structure of the phase diagram for several values of $d$-shell filling and various symmetry classes of lattice distortions.  Our results are relevant to the search for exotic topological insulators and quantum spin liquids in strongly correlated materials with strong spin-orbit coupling.
\end{abstract}
\date{\today}

\pacs{71.10.Fd,72.80.Ga,03.65.Vf}


\maketitle

\section{Introduction \label{introduction}}
Transition metal oxides have been an active topic of research for decades.\cite{Maekawa,Tokura_2000}
In particular, the interplay between Coulomb interaction, spin-orbit coupling, and lattice degrees of freedom have made transition metal oxides an ideal playground to test new theories and discover exotic behaviors. Notable among them are high temperature superconductivity,\cite{Lee:rmp06,Dagotto:rmp94} colossal magnetoresistance,\cite{Salamon:rmp01} and heavy fermion physics,\cite{Lee:prl01,Shimoyamada:prl06,Jonsson:prl07} among many other possiblities.\cite{Maekawa,Tokura_2000,Balents:nat10} The rather localized nature of $3d$ orbitals (compared to $4d$ and $5d$) in some transition metal oxides enhances the on-site electron-electron interaction and typically makes it a dominant energy scale.\cite{Maekawa,Tokura_2000} In the $3d$ transition metal oxides other interactions such as spin-orbit and electron-lattice coupling are typically small compared to the on-site Coulomb interaction and ground states with antiferromagnetic order are typical.\cite{imada:rmp98}

However, $4d$ and $5d$ orbitals in layered perovskites such as $\mathrm{Sr_2RuO_4}$, $\mathrm{Sr_2RhO_4}$, $\mathrm{Sr_2IrO_4}$, $\mathrm{Na_2IrO_3}$, and the hyperkagome $\mathrm{Na_4Ir_3O_8}$ are more spatially extended and thus the Coulomb interaction is typically weaker than those with $3d$ orbitals.\cite{ryden:prb70} The more extended nature of the $4d$ and $5d$ orbitals compared to the $3d$ orbitals leads to a greater level splitting in a crystal field, and enhances their sensitivity to lattice distortions. In many oxides, the transition ions are surrounded by an octahedron of oxygen atoms, $\mathrm{MO_6}$, where M represents a transition metal ion. The crystal field splits the 5 degenerate (neglecting spin for the moment) $d$-orbitals into two manifolds (see Fig.\ref{pyrochlore}c): a lower lying $t_{2g}$ ($d_{xy},d_{yz},d_{zx}$) manifold and a higher lying $e_{g}$ ($d_{3z^2-r^2},d_{x^2-y^2}$) manifold.\cite{Maekawa,Tokura_2000} The energy separation between the $t_{2g}$ and $e_{g}$ levels is conventionally denoted ``10Dq" and is typically on the order of  $\sim$1-4 eV, which is large compared to many $3d$ compounds.\cite{moon:prb06}

Besides the crystal field, the relativistic spin-orbit coupling is another energy scale that results from the large atomic numbers of heavy transition elements. While in the absence of spin-orbit coupling the on-site Coulomb interaction is of the same order as the band width,\cite{Nakatsuji:prl00,lee:prb01} inclusion of strong spin-orbit coupling modifies the relative energy scales.\cite{Pesin:np10} Thus, for materials with $4d$ and particularly $5d$ electrons, one expects the appearance of novel phases with unconventional electronic structure due to the characteristic energy of spin-orbit coupling approaching that of the Coulomb interactions.\cite{Pesin:np10} 

In a cubic environment, the $L=2$ orbital angular momentum of the $d$-orbitals is projected down to an effective angular momentum $l=1$ (with a minus sign) in the $t_{2g}$ manifold.\cite{Pesin:np10} When the spin-orbit coupling is also strong, neither spin nor orbital angular momentum is a good quantum number.  Instead, the total angular momentum, {\em i.e.} $j=l+s$, is a conserved quantity, where $s$ is the spin of the electron. Thus, spin-orbit coupling splits the $t_{2g}$ orbitals with spin into a $j=1/2$ doublet and $j=3/2$ quadruplet separated by an energy gap proportional to the strength of the spin-orbit coupling, $\lambda$.  (See Fig.\ref{pyrochlore}c.) For large enough spin-orbit coupling, the new effective spin states lead to a great deal of novel Mott insulating states,\cite{kim:prl08,BJkim:sci09,jin:prb09,singh:prb10,jin:arxiv09,watanabe:prl10} possible spin liquids in the hyper-kagome lattice,
\cite{okamoto:prl07,lawler:prl08,chen:prb08,zhou:prl08,lawler2:prl08,norman:prb10,micklitz:prb10,Bergholtz:prl10} orbital-oriented exchange coupling in Kitaev-type models,\cite{Jackeli:prl09,Chaloupka:prl10} Dirac semi-metal with Fermi arcs,\cite{wan:arxiv10} the quantum spin Hall effect,\cite{shitade:prl09} topological Mott insulators,\cite{Pesin:np10} topological magnetic insulators with axionic excitations,\cite{sczhang:np10,wang:arxiv10} and possibly high temperature superconductivity.\cite{fawang:arxiv10}

Of particular interest in this paper are the topological phases that occur in weak to moderately strongly interacting systems with strong spin-orbit coupling.\cite{Kane:prl05,Kane_2:prl05,Bernevig:prl06,Bernevig:sci06,Konig:sci07,Teo:prb08,Fu:prl07,Hsieh:nat08,Hsieh:sci09,Moore:prb07,Roy:06,Roy:prb09,Chen:sci09,Zhang:np09,Guo:prl09} (For excellent recent reviews see Refs.[\onlinecite{Moore:nat10},\onlinecite{hasan:rmp10}].) The possible time-reversal invariant topological phases of matter have non-trivial topological features in their global band structure and robust edge (surface) states.  When strong electron correlations are taken into account, spin-orbit coupling can give rise to a topological Mott insulating phase in which the charge degrees of freedom are completely gapped (even on the surface), but where the spin degrees of freedom inherit the non-trivial band topology from the weakly interacting limit.\cite{Pesin:np10,rachel:prb10,Young:prb08} Thus, the spin degrees of freedom form gapless spin-only edge (surface) modes. Such edge (surface) states can in principle be detected in thermal transport measurements (but not so readily in spin transport as the spin-orbit coupling generically destroys all spin conservation laws).

In this work we focus on the interplay and competition between strong correlation effects, spin-orbit coupling, and lattice distortion that is expected to be important in heavy transition metal oxides.   In the heavy transition metal oxides one expects both the spin-orbit coupling\cite{shitade:prl09,dodds:arxiv10} and the lattice distortion energies\cite{Landron:prb08} to be of the order of $0.05-0.5$ eV, while the interaction energy is typically at the higher end of this scale to somewhat larger, $0.5-2$ eV.\cite{shitade:prl09,wan:arxiv10}  While the phase diagram of an interacting undistorted pyrochlore model with $j=1/2$ has already been studied,\cite{Pesin:np10} we expand those results to include the effects of distortions of the local octahedra on the phase diagram. We also investigate pyrochlore oxides at different $d$-level fillings with the Fermi energy lying in the quadruplet $j=3/2$ manifold, which has not been considered in previous works. One of our motivations is to see if the $j=3/2$ manifold can also realize the interesting Mott phases of the $j=1/2$ manifold.\cite{Pesin:np10}  We find that, indeed, these exotic phases can be realized for the $j=3/2$ manifold.  Moreover, we find that for the $j=1/2$ manifold ``weak" topological variants of the exotic Mott phases can also appear in the phase diagram when certain types of lattice distortion are present.

This paper is organized as follows. In Sec. \ref{effective_H} we derive an effective nearest-neighbor tight-binding Hamiltonian that properly captures the non-interacting limit of the physics, and include an on-site Hubbard interaction term to describe electron correlations. In the absence of interactions and distortions, the ground state is metallic for weak spin-orbit coupling, and becomes a strong topological insulator as spin-orbit coupling grows. In Secs. \ref{Slave_rotor}, \ref{j12_band}, and \ref{j32_band} we study the effects of  Coulomb interaction and lattice distortion on equal footing using the slave-rotor mean field theory.\cite{florens:prb02,florens:prb04} As the strong correlation limit is approached, a Mott transition occurs and exotic phases are realized.  We discuss the conditions that favor these unusual phases. Finally,  we conclude in Sec.\ref{conclusions} and outline interesting topics for further study.

\section{Derivation of the effective Hamiltonian} \label{effective_H}

\begin{figure}[t]
\begin{center}
\includegraphics[width=8cm]{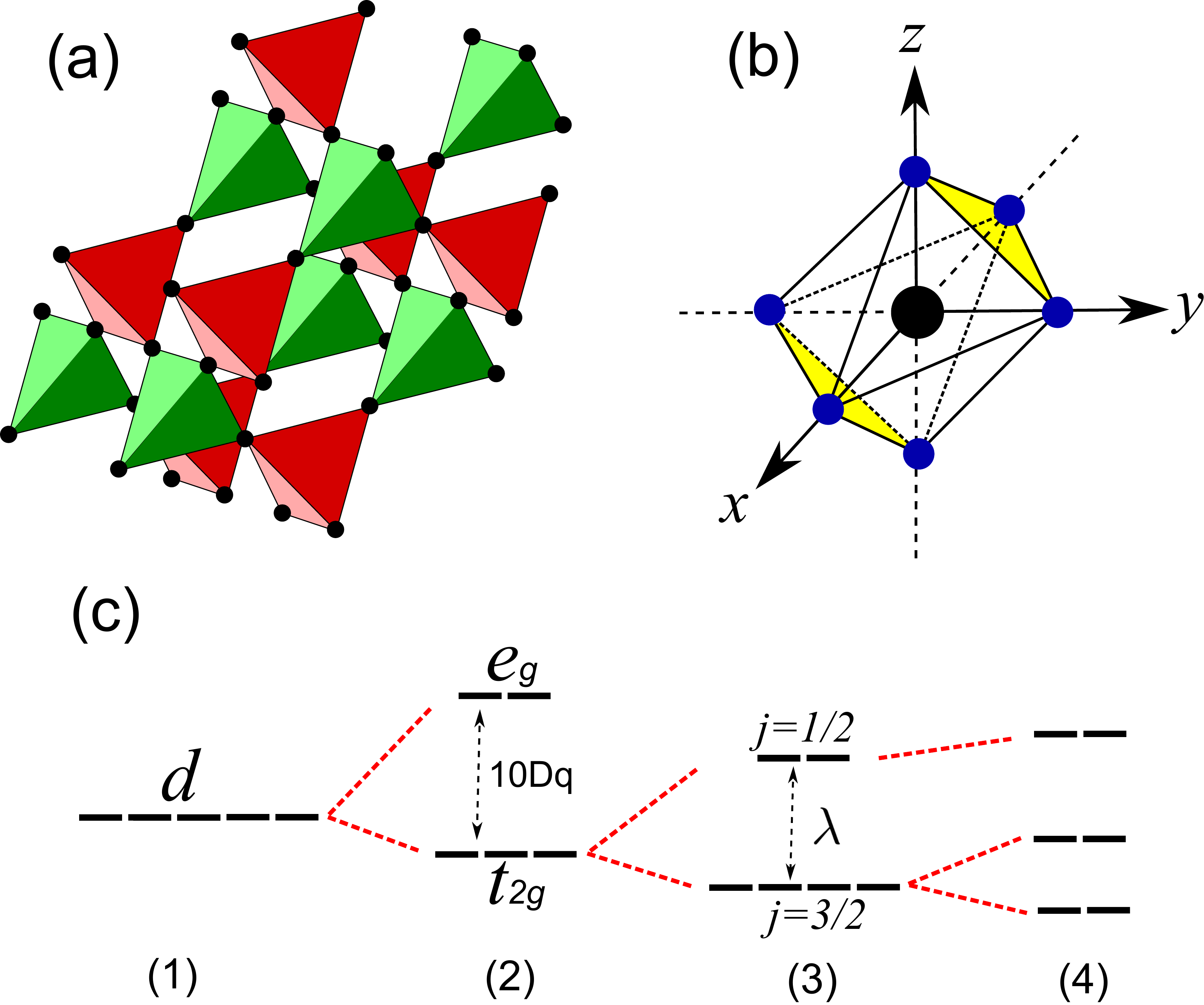}
\caption{(Color online) (a) An illustration of the pyrochlore lattice which is composed of corner sharing tetrahedra. Transition elements are indicated by black solid circles. (b) Each transition ion is surrounded by an oxygen octahedron shown by six solid blue (dark grey) circles. A transition ion is located at the origin of the local coordinate and is shown in black. We study a trigonal distortion  preserving $C_3$ symmetry applied along the [111] direction (or its equivalent), shown by two yellow (grey) faces, and an elongation preserving $C_4$ symmetry along the $z$-axis of the local coordinate. (c) A schematic representation of the splitting of the bare atomic $d$-levels (1), due to a cubic crystal field arising from the octahedral environment (2), unquenched spin-orbit coupling in the $t_{2g}$ manifold (3), and a distortion of the octahedron (4). The values of the splittings in (4) depend on $\lambda$ and $\Delta_{3,4}$.} \label{pyrochlore}
\end{center}
\end{figure}

In this work, we restrict our attention to pyrochlore oxides of the form $\mathrm{A_2M_2O_7}$, where A is a rare earth element and M is a transition metal element. Examples include  $\mathrm{A_2Ir_2O_7}$ (A=Y, Pr, Eu or other rear-earth elements), $\mathrm{Cd_2Os_2O_7}$, and $\mathrm{Cd_2Re_2O_7}$. In these materials, the transition metal elements form a pyrochlore lattice and each M sits in the center of an oxygen octahedron.\cite{yang:prb10}   The relevant geometry and coordinate system we use, along with the important level splittings, are shown in Fig.~\ref{pyrochlore}.

For strong spin-orbit coupling and with $d^{\alpha}$($\alpha \leq 4$, {\em i.e.}, less than 4 electrons) the upper $j=1/2$ manifold is empty and the important electronic structure is given by the lower $j=3/2$ manifold. In ordered double perovskites with a local quadruplet, {\em strong on-site interactions} add bi-quadratic and bi-cubic exchange interactions to the effective spin exchange Hamiltonian deep in the Mott regime.\cite{chen:arxiv10} It is also argued that the same local $C_3$ distortion that we consider in subsequent sections maps the exchange Hamiltonian into a pseduo-spin-$1/2$ model that favors magnetic order.\cite{dodds:arxiv10}  

Here we focus on the weak to moderately strongly interaction regimes.\cite{Guo:prl09,Pesin:np10,yang:prb10}  
We begin by deriving an effective tight-binding Hamiltonian between transition ions located at the vertices of the corner-shared tetrahedrons in the pyrochlore lattice shown in Fig.\ref{pyrochlore}a. The transition ions we consider have $5d$ orbitals with three or five electrons in the triply degenerate (neglecting spin) $t_{2g}$ manifold. The spin-orbit coupling in this manifold has the following form\cite{Maekawa}
\begin{equation} \label{local_so}
H_{so}=-\lambda l\cdot s,
\end{equation}
where $l=1$ and $s=1/2$ describe the orbital and spin degrees of freedom, and $\lambda>0$ parameterizes the strength of the spin-orbit coupling. That the $t_{2g}$ orbitals can be effectively described by angular momentum $l=1$ comes from the projection of the $d$-orbital angular momentum into the local basis of $t_{2g}$ manifold.\cite{Pesin:np10,chen:arxiv10}

To study the effects of lattice deformations,\cite{Bergmann:prb06} we assume that the octahedron surrounding an ion can be distorted in two ways: (1) a trigonal distortion preserving local $C_3$ symmetry and (2) an elongation (expansion) of octahedra preserving local $C_4$ symmetry. (See Fig.\ref{pyrochlore}b.) The former has been argued to be rather common and can be described by the following Hamiltonian on each transition metal ion site:\cite{yang:prb10} 
\begin{equation} 
\label{c3}
H_{tri}=-\Delta_3 (d_{yz}^{\dagger}d_{zx}+d_{yz}^{\dagger}d_{xy}+d_{zx}^{\dagger}d_{xy})+h.c.,
\end{equation}
where $\Delta_3$ parameterizes the strength and sign of the $C_3$ preserving distortion, and the $C_4$ elongation/contraction splitting is described by\cite{chen:arxiv10} 
\begin{equation}  
\label{c4} 
H_{el}=\Delta_4 l_{z}^{2}=\Delta_4(n_{yz}+n_{zx}), 
\end{equation} 
where $\Delta_4$ parametrizes the strength and sign of the distortion, and $l_z$ is the $z$ component of the effective angular momentum of the $t_{2g}$ orbitals related to the occupation of the $d_{xy}$ orbital by $n_{xy}=n_d-(l_z)^2$ which follows from the constraint $n_d=n_{xy}+n_{yz}+n_{zx}$.\cite{chen:arxiv10}  (See also Appendix \ref{app:L_z}.) For an elongation of the tetrahedron, $\Delta_4<0$, and for a compression of the tetrahedron, $\Delta_4>0$.  Trigonal distortions appear to be more common in real materials, and the magnitude of the energy splittings can be crudely estimated from density functional theory calculations based on X-ray determined positions of oxygen atoms around the transition metals.  We are not aware of detailed calculations of this type for the 4$d$ and 5$d$ pyrochlore oxides, but closely related 3$d$ systems appear to have splittings on the level of $0.01-0.5$ eV.\cite{Landron:prb08}  We take this as crude estimate, with the larger end of the energy scale probably more likely for the more extended 4$d$ and 5$d$ orbitals.

Thus, the local Hamiltonian describing the $t_{2g}$ orbitals on each site is
 \begin{equation} 
 \label{local}  
 H_{local}=H_{so}+H_{tri}+H_{el}.
 \end{equation}
The Hamiltonian \eqref{local} can be easily diagonalized and its eigenvectors describe a projection onto the spin-orbit plus distortion basis. We will denote the projection by a matrix $M$, which contains all the information about the spin-orbit coupling and the distortion of the octahedra (all assumed identical so translational invariance is preserved). Moreover, due to the presence of time-reversal symmetry, the eigenvectors form Kramers pairs. A schematic representation of splitting $t_{2g}$ upon including the terms in Eq.\eqref{local} is shown in Fig.\ref{pyrochlore}c.

We now turn to a derivation of the effective Hamiltonian. We first assume $\lambda=\Delta_3=\Delta_4=0$, {\em i.e.} neglect the contributions in \eqref{local}. To obtain the kinetic terms of the Hamiltonian, we need to describe the $t_{2g}$ orbitals of a single ion in the local coordinate system defined by the octahedron of oxygen atoms surrounded the ion, and we need the $p$-orbitals of oxygen in the global coordinate system.  The hopping of electrons from one transition metal ion to a nearest-neighbor transition metal ion is mediated by the oxygen $p$-orbitals.  (We note that for the relatively extended $5d$ orbitals direct overlap may also be important, as well as further neighbor hopping.\cite{jin:arxiv09}) We thus compute the $p$-$d$ overlaps to determine the hopping matrix elements. The local and global axes are related by a set of rotation matrices.\cite{Pesin:np10,yang:prb10} The combination of rotation matrices and $d$-$p$ overlaps gives rise to the following Hamiltonian:
\begin{equation} 
\label{H_d}
 H_{d}=\varepsilon_d \sum_{i\gamma\sigma}d_{i\gamma\sigma}^{\dag}d_{i\gamma\sigma}+t\sum_{<i\gamma \sigma,i'\gamma' \sigma'>}T^{ii'}_{\gamma \sigma,\gamma' \sigma'}d_{i\gamma\sigma}^{\dag}d_{i'\gamma'\sigma'},
 \end{equation} 
 where $i$, $\gamma$, and $\sigma$ in the sums run over lattice sites, $t_{2g}$ orbitals ($xy,yz,zx$), and spin degrees of freedom, respectively. The $\varepsilon_{d}$ stands for the onsite energy of the degenerate $t_{2g}$ orbitals, and $t=\frac{V^{2}_{pd\pi}}{\varepsilon_p-\varepsilon_d}$ is the unrotated hopping amplitude depending on the overlap integral $V_{pd\pi}$ and the energy difference between $p$ and $d$ orbitals. The parameter $t$ sets the basic hopping energy scale in the problem. Without loss of generality we set $\varepsilon_d=0$. 
 
 The effect of spin-orbit coupling and distortion are included via the projection of Hamiltonian in Eq.\eqref{H_d} into the eigenvectors of the local Hamiltonian in Eq.\eqref{local} using matrix $M$ as follows: 
 \begin{equation} 
 \label{H0} 
 H_{0}=\sum_{i\alpha}\upsilon_{\alpha}c^{\dag}_{i\alpha}c_{i\alpha}+t\sum_{<i\alpha,i'\alpha'>}\Gamma^{ii'}_{\alpha,\alpha'}c^{\dag}_{i\alpha}c_{i'\alpha'},
 \end{equation} 
 where $\upsilon_{\alpha}$ ($\alpha=1,...,6$) stands for the six eigenvalues of local Hamiltonian \eqref{local}, and the matrix $\Gamma$ describes the hopping between sites given in the local basis via $\Gamma=M^{*}TM^{T}$. The $c^{\dag}_{i\alpha}(c_{i\alpha})$ is the creation (annihilation) operator of an electron at site $i$ and in local state $\alpha$. Finally, we add a Coulomb interaction to obtain 
 \begin{equation}
 \label{hubbard}
 H=H_0+\frac{U}{2}\sum_{i}(\sum_{\alpha}c^{\dag}_{i\alpha}c_{i\alpha}-n_d)^2,
 \end{equation} 
 where $U$ is the on-site Coulomb interaction and $n_d$ is the number of electrons on the $5d$ orbital of the transition metal ion. In the remainder of this paper, we investigate the zero-temperature phase diagram of the full Hamiltonian \eqref{hubbard}, which includes the spin-orbit coupling and lattice distortions in \eqref{local}.

Before closing this section, it is instructive to take a look at the non-interacting limit of our model. In the absence of distortion, small spin-orbit coupling favors a metallic state for all fillings we consider (because they correspond to partially filled bands). However, strong spin-orbit coupling opens a gap at $\lambda/t=2.8$ for filling $n_d=5$,\cite{Pesin:np10} and for $\lambda/t=2.5$ at filling $n_d=3$ (see Fig.\ref{pd_undistortion}). The two fillings correspond to the $j=1/2$ and $j=3/2$ manifolds in the bands, with one hole/ion in each case. Therefore, strong spin-orbit coupling can turn the metallic band structure arising from the quadruplet $j=3/2$ manifold into a band insulating phase. 

By using the Fu and Kane\cite{Fu:prl07} construction, we can determine whether the insulating phase is trivial or topological. We make use of the inversion symmetry of the model and look for the parity of eigenstates at the time reversal invariant momenta. Those parities are related to the strong topological $\mathrm{Z_2}$ index $\nu_0$ via the following relation,
 \begin{equation} 
 \label{z2} 
 (-1)^{\nu_0}=\prod_{m}\prod_{i}\xi_{2m}(\Gamma_i), 
 \end{equation} 
 where the first product is taken over filled bands, the second one over the time reversal invariant momenta $\Gamma_i$, and $\xi_m(\Gamma_i)$ is the corresponding parity eigenvalue of band $m$ at time-reversal invariant momentum $\Gamma_i$. (Note that only one band from each set of Kramer's pairs is included in the product.  Note also, that we have followed convention and used $\Gamma_i$ with a single subscript for the time-reversal invariant momenta. It should not be confused with the multi-indexed $\Gamma^{ii}_{\alpha,\alpha'}$ in \eqref{H0} that describes the hopping.) The index $i$ is a collective index, {\em i.e.} $i\in \{i_1,i_2,i_3\}$, in terms of reciprocal lattice vectors $\mathrm{K}$ of the pyrochlore lattice: $\Gamma_i=1/2(i_1\mathrm{K_1}+i_2\mathrm{K_2}+i_3\mathrm{K_3})$ so that $i_1,i_2,i_3=0,1$. The week indices are defined in a similar way to the strong index: 
 \begin{equation}
 \label{weakz2} 
 (-1)^{\nu_j}=\prod_{m}\prod_{i(i_j=1)}\xi_{2m}(\Gamma_i), 
 \end{equation}
where $j=1,2,3$. Thus, the index $(\nu_0;\nu_1\nu_2\nu_3)$ defines sixteen classes of band insulators.  If $\nu_0=1$, the state is said to be a strong topological insulator (STI) and it has time-reversal symmetry protected gapless boundary excitations described by an odd number of Dirac cones in the surface state Brillouin zone.\cite{Fu:prl07}  On the other hand, if $\nu_0=0$ but $\nu_j\neq 0$ for at least one $j\in (1,2,3)$ then the state is said to be a weak topological insulator (WTI).  In this case, gapless surface modes may be present in a clean system, but can be destroyed with disorder. As we will see, we find both STI and WTI (including strong correlation generalizations) in our model when lattice distortions are present. (See Fig.\ref{j12_ph} lower panel.)

\section{slave-rotor mean field theory and self-consistent equations }
\label{Slave_rotor}
In this section we apply the slave-rotor mean-field theory developed by Florens and Georges\cite{florens:prb02,florens:prb04} to treat the effect of weak to intermediate strength Coulomb interactions in the regime where the charge fluctuations remain important. In this theory each electron operator is represented in terms of a collective phase, conjugate to charge, called a rotor and an auxiliary fermion called a spinon as
\bea 
\label{slave_rotor} 
c_{i\alpha}=e^{i\theta_i}f_{i\alpha}, 
\eea 
where $c_{i\alpha}$ is the electron destruction operator at site $i$ with quantum number $\alpha$, representing the states in \eqref{hubbard}. The factor $e^{i\theta_i}$ acting on the charge sector is a rotor lowering operator (with $\theta_i$ a bosonic field), and $f_{i\alpha}$ is the fermionic spinon operator.  The product of the two results in an object with fermi statistics, need for the electron. Note the rotor part only carries the charge degree of freedom while the spinon part carries the remaining degrees of freedom $\alpha$. Therefore, an electron has natural spin-charge separation if $\alpha$ is spin in this representation. 

A constraint should be imposed to retain the physical states as
\begin{equation}
\label{constraint}
L_i+\sum_{\alpha}f_{i\alpha}^{\dag}f_{i\alpha}=n_d,
\end{equation}
where $L_i$ is number of rotors. Using this representation, the interacting Hamiltonian \eqref{hubbard} can be written as,
\begin{eqnarray}
\label{H_slave}
H=\sum_{i\alpha}(\upsilon_{\alpha}-\mu)f_{i\alpha}^{\dag}f_{i\alpha}+t\sum_{<i,i'>}\Gamma^{ii'}_{\alpha,\alpha'}e^{-i(\theta_i-\theta_{i'})}f^{\dag}_{i\alpha}f_{i'\alpha'}\nonumber\\
+\frac{U}{2} \sum_i L_i^2, \hspace{1.7cm}
\end{eqnarray}
where a chemical potential $\mu$ has been introduced. In order to treat the phase $\theta$ and angular momentum $L$ on an equal footing, we need to switch from ($\theta,L$) to fields ($\theta,i\partial_\theta$)\cite{florens:prb04} so that $i\partial_{\tau}\theta=\frac{\partial H}{\partial L}$, which gives $L=(i/U)\partial_\tau \theta$. The corresponding action is 
\begin{equation}
S=\int_{0}^{\beta}d\tau[-iL\partial_{\tau}\theta+H+f^{\dag}\partial_{\tau}f].
\end{equation}
We next introduce bosonic operators $X_i=e^{i\theta_i}$ and recast the action into
\begin{widetext}
\begin{equation} 
\label{action}
S=\int_{0}^{\beta}d\tau\bigg[\sum_{i\alpha}f_{i\alpha}^{\dag}(\partial_{\tau}+\upsilon_{\alpha}-\mu-h_i)f_{i\alpha}+\sum_{i}[\frac{1}{2U} \partial_{\tau}X_{i}^{*}\partial_{\tau}X_{i}+\frac{h_i}{2U}(X_{i}^{*}\partial_{\tau}X_{i}-c.c)+\rho_i (|X_{i}|^{2}-1)]+t\sum_{<i,i'>}\Gamma^{ii'}_{\alpha,\alpha'}X^{*}_{i}X_{i'}f^{\dag}_{i\alpha}f_{i'\alpha'}\bigg], \end{equation}
\end{widetext}
where $h_i$ and $\rho_i$ are Lagrange multipliers imposing the constraints $L_i+\sum_{\sigma}f_{i\sigma}^{\dag}f_{i\alpha}=n_d$ and $|X_i|^{2}=1$, respectively on each site. We have effectively carried out the integration over $L$ by using the relation $L=(i/U)\partial_\tau \theta$. The action \eqref{action} describes the coupled spinon and rotor degrees of freedom. We will assume that translational symmetry is preserved and decompose \eqref{action} into two parts by use of the following mean-field ansatz:
\begin{eqnarray}
\label{ansatz}
Q_f=\langle X_{i}^{*}X_{i'}\rangle,
\nonumber \\
Q_{\theta}= \sum_{\alpha \alpha'}\Gamma^{ii'}_{\alpha,\alpha'}\langle f_{i\alpha}^{\dag}f_{i'\alpha'}\rangle.
\end{eqnarray}
Then, the action in Eq.(\ref{action}) can be written as $S=S_f+S_{\theta}$, in which
\begin{widetext}
\bea
\label{action_f} &&S_f=\int_{0}^{\beta}d\tau\bigg[\sum_{i\alpha}f_{i\alpha}^{\dag}(\partial_{\tau}+\upsilon_{\alpha}-\mu-h)f_{i\alpha}+ tQ_{f}\sum_{<i,i'>}\Gamma^{ii'}_{\alpha,\alpha'}f^{\dag}_{i\alpha}f_{i'\alpha'}\bigg],
\\ &&S_{\theta}=\int_{0}^{\beta}d\tau\bigg[\sum_{i}[\frac{1}{2U}\partial_{\tau}X_{i}^{*}\partial_{\tau}X_{i}+\frac{h}{2U}(X_{i}^{*}\partial_{\tau}X_{i}-c.c)+\rho (|X_{i}|^{2}-1)]+tQ_{\theta}\sum_{<i,i'>}X^{*}_{i}X_{i'}\bigg].
\eea
\end{widetext}

The chief benefit of the above actions $S_f$ and $S_{\theta}$ is that they are quadratic in spinon and rotor fields, and therefore the calculation of the corresponding Green's function is straightforward. One can simply use Fourier transformation and go to the eigenfunction basis to obtain
\bea \label{fouri}&&
S_f=\sum_{k,n,j}[\widetilde{f}_{knj}^{\dag}(i\omega_{n}+\epsilon_{j}(k))\widetilde{f}_{knj}],
\eea
where $k$ is the momentum, $j$ labels the four sites in a unit cell as well as the effective spin degrees of freedom, and the dispersion of band $j$ is given by $\epsilon_{j}(k)$. The $i\omega_{n}$ are fermionic Matsubara frequencies and $\widetilde{f}_{knj}$ is a linear combination of $f_{k\alpha}$ that diagonalizes the spinon part of the Hamiltonian:
\bea
\label{spinonH} H_f=\sum_{i\alpha}f_{i\alpha}^{\dag}(\upsilon_{\alpha}-\mu)f_{i\alpha}+ tQ_{f}\sum_{<i,i'>}\Gamma^{ii'}_{\alpha,\alpha'}f^{\dag}_{i\alpha}f_{i'\alpha'}.\nonumber \\ \eea 
The rotor action reads
\bea
\label{rotorH} S_{\theta}=\sum_{k,n,j}\left[\widetilde{X}_j^{*}(k,\nu_n)\left[\frac{\nu_{n}^{2}}{U}+\rho+tQ_{\theta}\xi(k)\right]\widetilde{X}_j(k,\nu_n)\right]\nonumber,
\\
\eea
where the parameter $U$ has been replaced by $U/2$, so that the atomic limit is treated correctly.\cite{florens:prb02} Note that we have set $h=0$, which guaranties that the constraint Eq.(\ref{constraint}) is satisfied on the mean-field level.\cite{florens:prb02} The $\nu_{n}$ are bosonic Matsubara frequencies and $\xi(k)$ is related to the spectrum of the rotor Hamiltonian\cite{Pesin:np10} via  $\xi_{1,2}(k)=2\left(1\pm\sqrt{1+\sum_{\mu<\nu}\cos(k_{\mu}/2)\cos(k_{\nu}/2)}\right)$, and $\xi_{3,4}=-2.$  In the sum under the radical, $\mu,\nu$ run over three values and give max$\{|\xi_i|\}$=6.

The spinon and rotor Green's function can be readily written as
\bea
 \label{green} 
G_{f}(k,i\omega_n)=\frac{1}{i\omega_{n}+\epsilon(k)},\\
G_{\theta}(k,i\nu)=\frac{1}{\frac{\nu_{n}^{2}}{U}+\rho+tQ_{\theta}\xi(k)}, 
\eea
which can be used to to determine the character of spinon and rotor excitations.  We note that the spinons are single-particle-like excitations, while the rotors represent collective excitations.
 
The self-consistent equations that should be solved to determine the phase boundaries in the phase diagram are (where the rotors are assumed condensed)
\bea
\label{self1}
1=|X_i|^{2}=\frac{1}{4N}\frac{1}{\beta}\sum_{n,j,k}\frac{1}{\frac{\nu_{n}^{2}}{U}+\rho+tQ_{\theta}\xi_{j}(k)},\\
Q_{f}=-\frac{1}{24N}\frac{1}{\beta}\sum_{n,j,k}\frac{\xi_{j}(k)}{\frac{\nu_{n}^{2}}{U}+\rho+tQ_{\theta}\xi_{j}(k)},\\
Q_{\theta}= \sum_{\alpha \alpha'}\Gamma^{ii'}_{\alpha,\alpha'}\langle f_{i\alpha}^{\dag}f_{i'\alpha'}\rangle, \\
4n_{d}=\frac{1}{N}\sum_{j,k}\Theta(\mu-\epsilon_{j}(k)),
\eea
where $N$ is the number of unit cells in the lattice, $\beta$ is the inverse temperature, and $\Theta(x)$ is the step function.
By use of the Matsubara sum\cite{rachel:prb10} $1/\beta\sum_{n}G_{\theta}(k,i\nu_{n})=\frac{U}{2\sqrt{U(\rho+\xi(k))}}$, the first two equations above in the set starting with \eqref{self1} can be simplified as\cite{Pesin:np10}
\bea 
\label{self21}
1=\frac{1}{4N}\sum_{j,k}\sqrt{\frac{U}{4(\rho+tQ_{\theta}\xi_{j}(k))}}\;,\\ \label{self22}
Q_{f}=-\frac{1}{24N}\sum_{j,k}\xi_{j}(k)\sqrt{\frac{U}{4(\rho+tQ_{\theta}\xi_{j}(k))}}\;.
 \eea

The rotor condensed phase is characterized by a nonzero value of $Z\equiv \langle e^{i\theta}\rangle$, where the electron operator is proportional to the spinon operator: $c_{i\alpha}=Zf_{i\alpha}$. The condition for condensation is that the gap of the rotor's spectrum $\Delta_{g}=2\sqrt{U(\rho+6tQ_{\theta})}$ closes. Therefore, right at the phase boundary $\rho=-6tQ_{\theta}$. Combined with equations Eq.\eqref{self21} and Eq.\eqref{self22}, $Q_{f}$ and the critical $U$ can be determined.\cite{Pesin:np10,rachel:prb10}

\section{$j=1/2$-band model: Effect of trigonal distortion}
\label{j12_band}

In this section we discuss the slave-rotor mean-field phase diagram of the $j=1/2$ model studied in Ref.[\onlinecite{Pesin:np10}] when a $C_3$ symmetry preserving trigonal distortion of the oxygen octahedra is included.\cite{yang:prb10} The effect of the trigonal distortion on the atomic $t_{2g}$ levels is given by  Eq.\eqref{c3}, which describes the compression or expansion of the octahedra along the [111] direction or any equivalent direction in the local coordinate.  Fig. \ref{pyrochlore}b shows a schematic of this deformation as indicated by the shaded faces of the octagon.
This distortion splits the $t_{2g}$ manifold into a singlet $|a\rangle$ with energy $\varepsilon_{a}=-2\Delta_3$ and doublet $|e_{g}'\rangle$ with energy $\varepsilon_{e}=\Delta_3$. 

The effect of spin-orbit coupling on the doublet can be understood by noting that the spin-orbit coupling in the subspace spanned by the doublet states acts like a Zeeman field,\cite{jin:arxiv09} \bea 
\label{c3_zeeman}
\langle H_{so}\rangle_{e_{g}'}=\frac{\lambda}{2}\tau^{z}\otimes \vec{\sigma}\cdot \vec{n},
\eea 
where $\tau^z$ and $\vec{\sigma}$ act on the pseudospin space spanned by $e_{g}'$ states and real spin, respectively. The unit vector $\vec{n}$ points in the direction of the trigonal distortion. Thus a gap may open by tuning the spin-orbit coupling in the presence of a strong distortion, which is consistent with density functional calculations for the iridate $\mathrm{Na_{2}IrO_{3}}$.\cite{jin:arxiv09}

\begin{figure}
\begin{center}
\includegraphics[width=8cm]{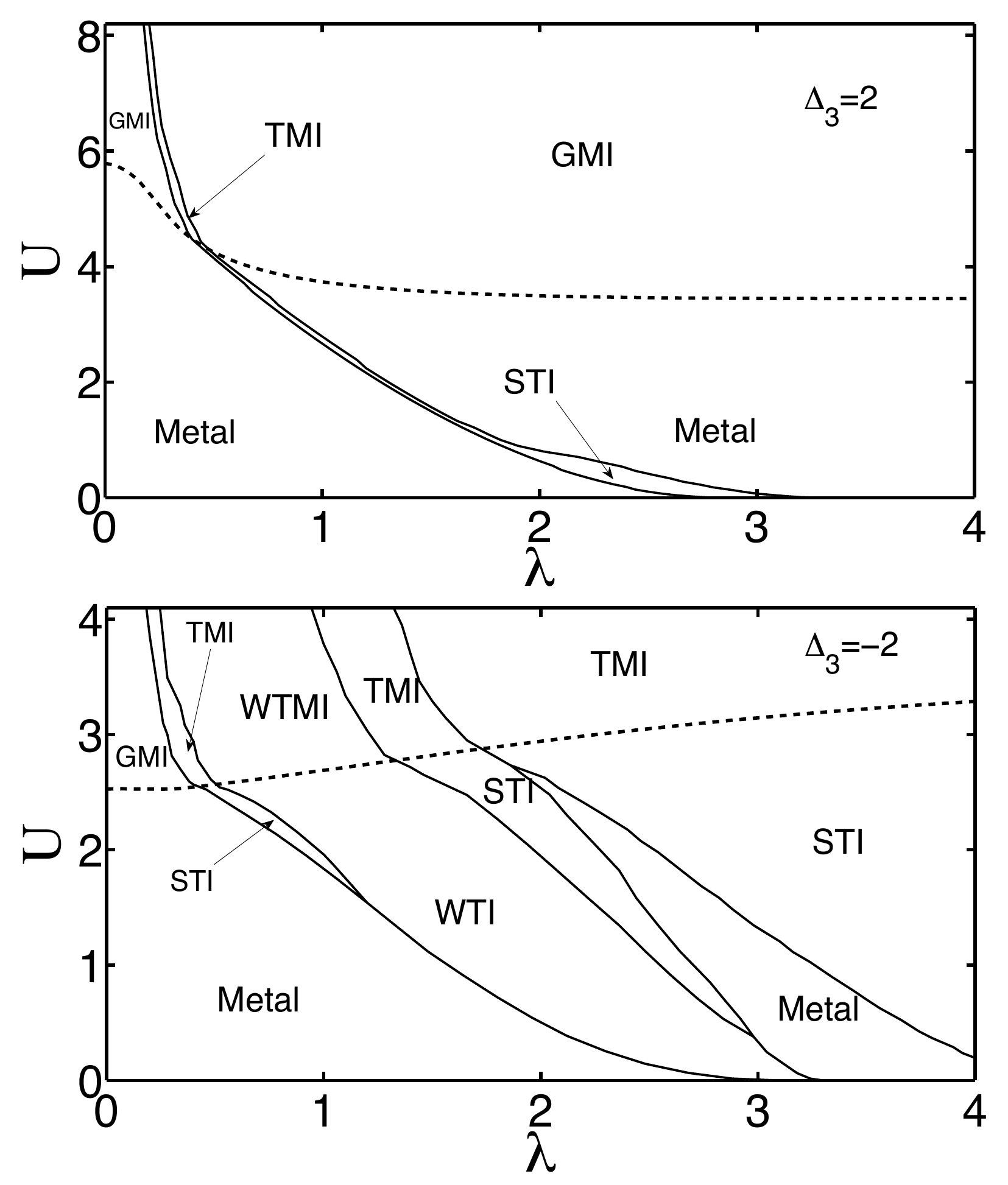}
\caption{Phase diagram of the $j_{eff}=1/2$-band model corresponding to $n_{d}=5$ with positive $\Delta_3=2t$ (upper panel) and negative trigonal distortion $\Delta_3=-2t$ (lower panel). The dashed line separates the rotor condensed phases (below) from the uncondensed phases (above). We set $t=1$, and the phases labeled are as follows: Strong topological insulator (STI), Weak topological insulator (WTI), Gappless Mott insulator (GMI), Topological Mott insulator (TMI), Weak topological Mott insulator (WTMI) and Metallic phases.} \label{j12_ph}
\end{center}
\end{figure}

As discussed in Ref.[\onlinecite{yang:prb10}], in the non-interacting limit a strong trigonal distortion can turn a strong topological insulator into a metal.  We would like to understand to what degree this happens in the presence of interactions, and to what degree the undistorted, interacting phase diagram of Pesin and Balents\cite{Pesin:np10} is changed by distortions. We thus study both interaction and distortion on equal footing. After verifying that our calculations successfully reproduce the phase diagram in Ref.[\onlinecite{Pesin:np10}], we first consider the case of $\Delta_3>0$. Our results are shown in Fig.\ref{j12_ph}. The upper panel exhibits the phase diagram with positive distortion. The thick dashed line separates the rotor condensed phase (below the line) from the uncondensed part (above the line). At the non-interacting level and for weak spin-orbit coupling, the system is in the metallic phase. However, a small window of gap opening exists for $\lambda \approx 2.8t-3.3t$, which is a strong topological insulating (STI) phase with $Z_2$ invariant (1;000). This small window forms a narrow gapped region in the phase diagram along the $U=0$ axis.  Note the ``re-entrant" metallic phase in the presence of distortion is different from the robust insulating phase found in Ref.[\onlinecite{Pesin:np10}] for large $\lambda$, and is qualitatively similar to the non-interacting distortion results found in Ref. [\onlinecite{yang:prb10}].
 
Within the slave-rotor mean-field theory, one finds that a narrow window of STI persists to interactions of order the bandwidth, after which it becomes a tiny sliver.  A metallic phase has mostly replaced what would be the STI in the absence of distortions.  As interactions are further increased, and a Mott transition occurs to gap out the rotor degree of freedom (above the dashed line), one finds the metal is converted into a gapless Mott insulator (GMI), which is a type of spin liquid with gapless bulk spin excitations\cite{Pesin:np10} described by  the spinon Hamiltonian in Eq.\eqref{spinonH}.  The only effect of the interaction $U$ is to renormalizes $Q_{f}$ through the self-consistent equations. The large regions of GMI indicates that lattice distortions of $\Delta_3>0$ type may be helpful in the realization of a gapless spin liquid state in this class of materials.  A very tiny sliver of STI is converted into a topological Mott insulator (TMI) above the line for which the rotors are no longer condensed.\cite{Pesin:np10}  Thus, distortions  of this type are detrimental to the realization of the TMI phase and suggest this phase may not be stable against lattice distortions.  

Next we consider the case of $\Delta_3<0$. The phase diagram is shown in the lower panel of Fig.\ref{j12_ph}. One distinctive feature compared to the case of positive distortion is that the critical $U_c$ for the Mott transition (dashed line) grows as the spin-orbit coupling strength is increased. A second feature is that a variety of phases appear from the interplay of spin-orbit coupling and correlation effects. At zero interaction, while small and intermediate values of spin-orbit coupling favor the metallic phase, a gap is opened for $\lambda \approx 3.1t-3.3t$. According to the $Z_2$ classification\cite{Kane_2:prl05} this insulating phase is a weak topological insulator (WTI) with $Z_2$ invariant (0;010), and the small window survives and persists up to intermediate interactions. To the best of our knowledge, this is the first identification of a weak topological insulator in an interacting model.  Gapless modes along certain classes of defects may be a way to identify this state in experiment.\cite{Ran:np09}

Another interesting feature of the phase diagram for $\Delta_3<0$ is that the correlation effect can drive the metallic and weak topological insulator phases into a strong topological insulating phase. All realized phases in the condensed rotor phase (below the dashed line) have carriers with both spin and charge. When the correlation is strong enough to strip the charge degree of freedom (above the dashed line), those phases will turn into the corresponding phases with only spin degrees of freedom. As before, the metallic and strong topological insulator phases are transformed into the GMI and TMI phases, respectively. Moreover, the weak topological insulator phases realize a novel the weak topological Mott insulator (WTMI) with increased interactions. These latter phases are absent on the undistorted lattice\cite{Pesin:np10} and to the best of our knowledge, is the first time the WTMI phase has been identified in a calculation.  It will have gapless thermal transport along the same class of defects that would have gapless charge (and thermal) transport in the WTI phase.\cite{Ran:np09} We note that there is an accidental gap closing in the TMI phase where two TMI phases are separated by a boundary. However, since the gap closing occurs at an even number of Dirac-like nodes, the topological properties remain unchanged through the gap closing points.

\section{$j=3/2$-band model: phase diagram of undistorted lattice}
\label{j32_band}
To date, the search for time-reversal invariant insulators in transition metal oxides has primarily focused on the $j=1/2$ manifold because of its obvious connection to the $s=1/2$ manifold heavily studied in the theoretical literature thus far.\cite{kim:prl08,shitade:prl09,Pesin:np10,yang:prb10,wan:arxiv10}
While the $j=1/2$ manifold is relevant for 1/2-filling ($n_d=5$), the $j=3/2$ manifold is relevant for $n_d=2$ which occurs in $\mathrm{Cd_{2}Re_{2}O_{7}}$ and $n_d=3$ which occurs in $\mathrm{Cd_{2}Os_{2}O_{7}}$.\cite{mandrus:prb01,singh:prb02}  In this section we investigate whether topological phases are still possible in the $j=3/2$ manifold for some range of $\lambda$ and study the phase diagram in the presence interactions, as was done for $j=1/2$ in Ref.[\onlinecite{Pesin:np10}].  In the next section, we will consider the effects of distortion on the $j=3/2$ phase diagram.

\begin{figure}
\begin{center}
\includegraphics[width=7.7cm]{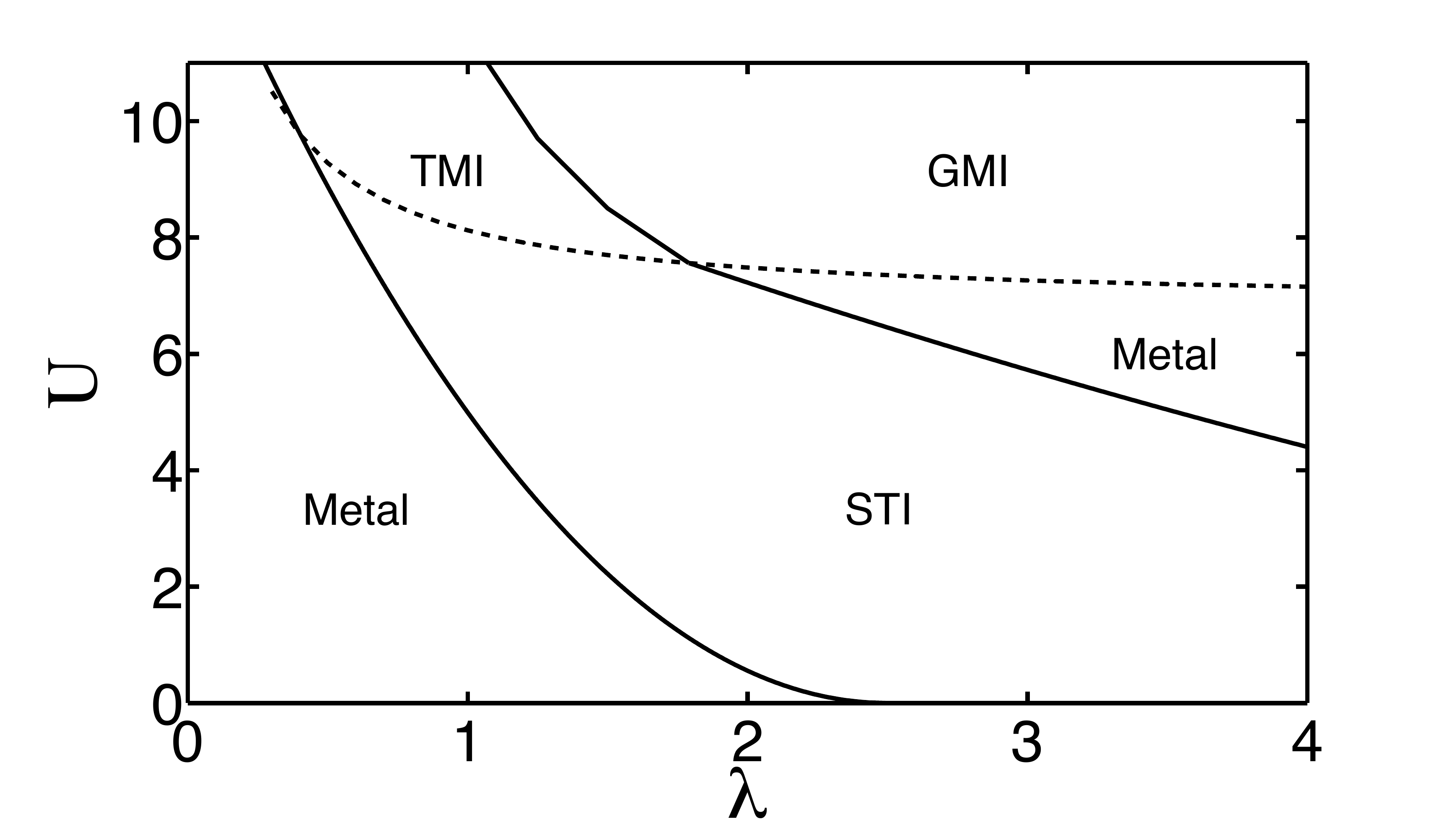}
\caption{Phase diagram of $j=3/2$-band model on the undistorted lattice for $n_d=3$. The abbreviations used are the same as those in Fig.\ref{j12_ph}. The dashed line separates the rotor condensed phase (below) from rotor uncondensed phase (above). All energies are expressed in units of $t$, as before.  Compared to the corresponding phase diagram for $j=1/2$ with $n_d$=5, the STI and TMI occupy a much smaller portion of the phase diagram.\cite{Pesin:np10} The phase diagram for $j=2/3$ with $n_d=2$ has no topologically non-trivial phases within our model, even in the presence of distortion.} \label{pd_undistortion}
\end{center}
\end{figure}

In our calculations, we find that the non-interacting model with $5d^{2}$ remains metallic for all physical spin-orbit coupling, though we can open a gap by distortion. A direct evaluation of the $Z_2$ invariant shows that the distortion-induced insulating phase is a trivial insulator. So we will focus on the case with a $5d^{3}$ electron configuration, with $\mathrm{Cd_{2}Os_{2}O_{7}}$ one possible example.\cite{mandrus:prb01,singh:prb02} In the non-interacting limit and for small values of spin-orbit coupling, we find a metallic phase.  However, for $\lambda \approx 2.5t$ a gap opens and a STI appears. The STI phase is characterized by the $Z_2$ indices $(1;000)$ and survives to moderate interaction. The corresponding phase diagram is shown in Fig. \ref{pd_undistortion}. The metallic phase is still present in a large portion of the phase diagram if the spin-orbit coupling is not too strong. It results in part from the fact that corresponding band structure coming from the $j=3/2$ manifold has a larger band width than the upper $j=1/2$ manifold and therefore has a weaker correlation effect for the same value of $U$.

The phase boundary that separates the STI phase from the weakly interacting metallic phase can be determined analytically.\cite{Pesin:np10} One notes that in the noninteracting limit a gap is opened at $\lambda \approx 2.5t$. Therefore, the $\lambda$ and $U$ that respect $\lambda \approx 2.5 Q_{f}(U)$ form the critical line between the STI and the metal. As in the case of the $j=1/2$ manifold, the Mott transition between the metallic (STI) phase and the GMI (TMI) phase is characterized by a vanishing amplitude of the rotor condensate at some critical value $U_c(\lambda)$ (given by the dashed line). The GMI-metal transition and its extension to finite temperature, which appears to be a possible Mott transition between a spin-liquid insulator and metal, has already been studied with a possible connection to the experimental observations in $\mathrm{Na_4Ir_3O_8}$.\cite{podolsky:prl09,lawler2:prl08,zhou:prl08}

The TMI phase is described by a gapped bulk spectrum of the spinon Hamiltonian, and is in the same topological phase appearing in the $j=1/2$ manifold.\cite{Pesin:np10} It supports gapless surface states of charge-neutral spinons. The same spin-charge separation also occurs in 2D cases,\cite{rachel:prb10,Young:prb08} where a quantum spin Hall state turns into an exotic quantum spin Hall effect at an intermediate regime of on-site Hubbard interaction. However, the 2D nature of the phase suffers from an instability due to fluctuations of the gauge field.\cite{rachel:prb10,Young:prb08}  

An analysis of symmetries reveals that the Hubbard model in Eq.\eqref{hubbard} has a $U(1)$ gauge symmetry as the slave-rotor representation of the physical electron in Eq.\eqref{slave_rotor} is invariant under the following gauge transformation: $f_{i\alpha}\rightarrow e^{i\varphi_{i}}f_{i\alpha}$ and $\theta_{i}\rightarrow\theta_{i}-\varphi_{i}$.  In the insulating exotic state, the rotors can be integrated out since the charges are gapped. The resulting theory is a compact $U(1)$ gauge theory coupled to the spinons.\cite{lee:prl05} The later theory is not stable against the fluctuations of the gauge field as it is a confining compact theory in 2D.\cite{polyakov:pyslett75} Such confinement renders the states unstable in 2D as the  gauge fields confine the free spinon-like excitations, effectively removing them as legitimate low-energy excitations.\cite{polyakov:pyslett75,herbut:prb03,herbut:prl03} While it is believed that the extension of the spin index to $N$ flavors renders it deconfining for sufficiently large $N$, the value of the critical $N$ is not known.\cite{hermele:prb04}  In order to stabilize the edge modes, the gauge fluctuations must be screened by other gapless degrees of freedom. In the 2D case, this can be done by use of a bilayer structure in which the ``second" layer contains the necessary gapless degrees of freedom.\cite{Young:prb08}  In spite of the shortcomings of the slave-rotor mean-field theory in 2D, we note that recent quantum Monte Carlo calculations on the Kane-Mele-Hubbard model show a similar phenomenology in some respects at intermediate interaction strength.\cite{Hohenadler:arxiv10,Zheng:arxiv10,Yamaji:arxiv10}

Returning to 3D, we note that in the TMI phase the gapless spinon surface states are coupled to the bulk 3D gauge fields.\cite{Witczak:prb10}  Thus, the low energy theory of the TMI phase is given by the spinon surface states coupled to the 3D gauge fields.  This theory is believed to be stable,\cite{Pesin:np10} as the gauge propagator is suppressed so that the spinons become better defined (the self-energy scales as the energy itself, up to logarithmic corrections).\cite{Witczak:prb10} Thus, the lowest order calculation in the U(1) gauge fluctuations suggests they are marginal; a more careful scaling analysis suggest they are actually marginally irrelevant.\cite{Witczak:prb10}  

Unlike the STI phase, the surface states of the TMI phase can not be characterized by electrical transport measurements due to the charge neutrality of the spinons on the surface of the TMI. Moreover, because this neutrality, there are no Friedel oscillations around a charged impurity on the surface.  However, spinon surface states of the TMI can be detected in thermal measurements, and by the way in which they modify the RKKY interaction between magnetic impurities at the surface.\cite{Witczak:prb10} In the GMI phase, on the other hand, the bulk specific heat behaves\cite{podolsky:prl09} as $C\sim T \ln(1/T)$, while in the metallic state it behaves as $C\sim T$.

\section{$j=3/2$-band model: effect of distortions}
\label{j32_distorion}
Having obtained the phase diagram of the undistorted $j=3/2$ model in Fig.\ref{pd_undistortion}, 
we now study the effect of the local distortion of octahedra introduced in Sec.\ref{effective_H}. We focus on two kinds of distortion with different symmetries: (1) a trigonal distortion of oxygen octahedra that preserves the $C_3$ symmetry, and (2) a compression and elongation of the oxygen octahedra that preserves the $C_4$ symmetry.  We use parameter $\Delta_3$ to describe the $C_3$ distortion and $\Delta_4$ to describe the $C_4$ distortions. The relevant Hamiltonians are given in \eqref{c3} and \eqref{c4}.  

We restrict our attention to the weak and intermediate interaction limit, so we neglect possible magnetic phases that could become favorable in the strong correlation limit. In that limit geometrical distortions could alter the isotropic antiferromagnetic superexchange and the combined effects of spin-orbit coupling and distortion can give rise to an anisotropic pseudo-spin Heisenberg model for some perovskites.\cite{dodds:arxiv10}

\begin{figure}[t]
\begin{center}
\includegraphics[width=7.7cm]{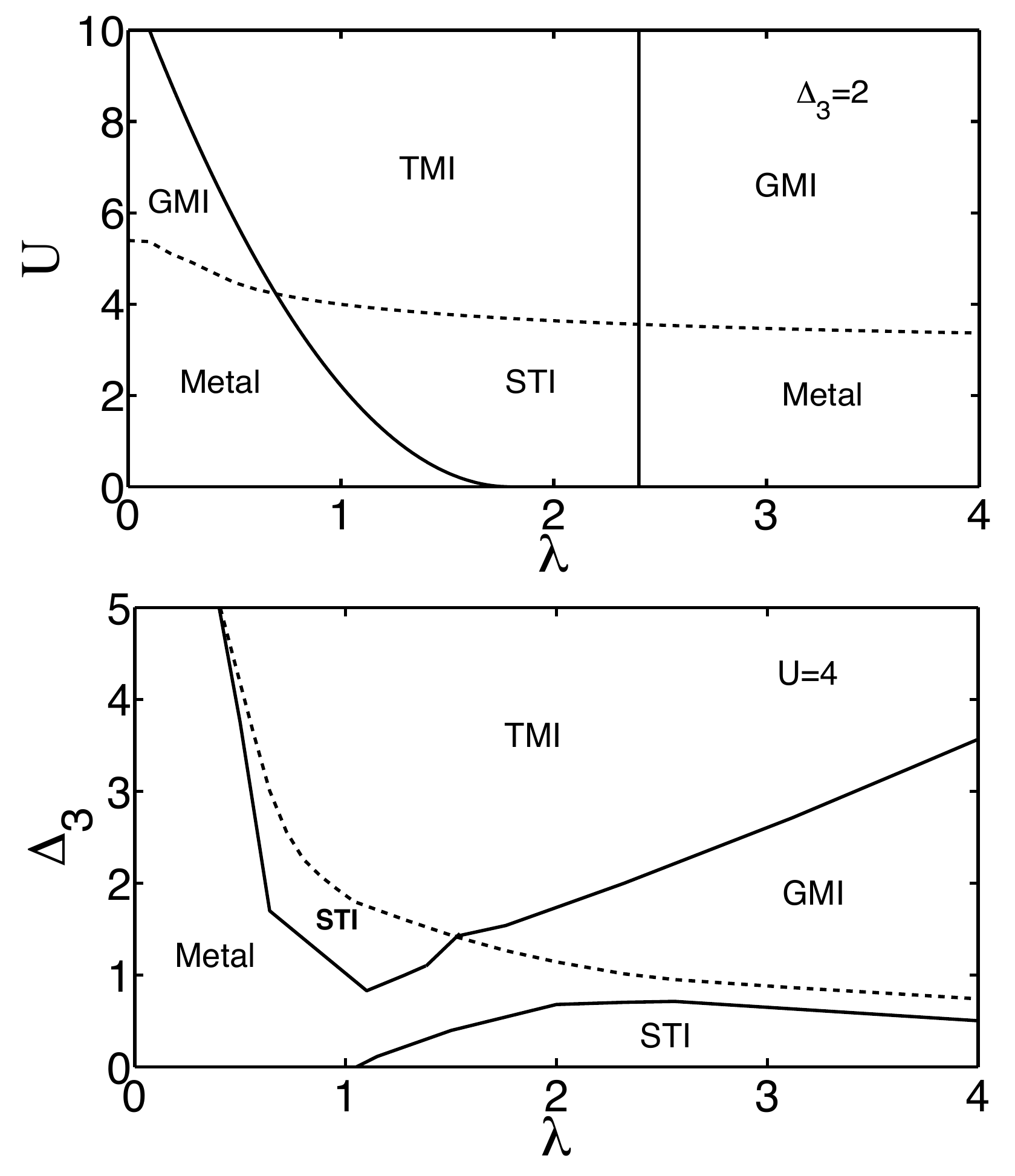}
\caption{Phase diagram of the $j=3/2$ band model with $n_d=3$, including the trigonal distortion of the octahedra. The labeling of the phases is the same as that used in Fig.\ref{j12_ph}. In the upper panel  $\Delta_3=2t$.  In the lower panel the interaction is fixed at $U=4t$ and the strength $\Delta_3$ of the trigonal distortion is varied, illustrating possible phases that may arise upon the application of pressure to a real system.  All energies are expressed in units of $t$. In both phase diagrams the dashed line separates the rotor uncondensed phase (above) from the condensed phase (below). We note the ``pocket" of STI around $\lambda \approx 1,\Delta_3\approx 1.5$ in the lower figure has a numerically difficult to determine boundary with the metallic phase; we have present our best assessment. } \label{j32_c3}
\end{center}
\end{figure}

\subsection{Trigonal Distortion of Oxygen Octahedra}
We first consider the trigonal distortion in Eq.\eqref{c3} on the $j=3/2$, $n_d=3$ manifold. The resulting phase diagrams for fixed $\Delta_3$ and fixed $U$ are shown in Fig.\ref{j32_c3}. We find the general structure of the phase diagram depends on $\Delta_3$ in a complicated way.  As one example, in the upper panel of Fig.\ref{j32_c3} we show the case of fixed trigonal distortion $\Delta_3=2t$. In that case, we find that the non-interacting model is dominated by the metallic phase, which is analogous to the trigonal distortion driven metallic phase on the $j=1/2$-band model, as discussed in Sec.\ref{j12_band} and also in Ref.[\onlinecite{yang:prb10}]. The gap opens at $\lambda \approx 1.8t$ and closes at $\lambda \approx 2.4t$. Therefore, in the presence of trigonal distortion only a small window of $\lambda$ admits the STI phase in the non-interacting limit. However, this small window grows with small but increasing interaction strength which helps to stabilize the STI phase.\cite{Pesin:np10} The metallic phase, however, remains dominant for $\lambda \gtrsim 2.4t$, even in the presence of interaction. We note that the boundary separating STI phase from the metallic phase around $\lambda \approx 2.4t$ is not exactly but very close to a straight line.

From the phase diagrams shown in Fig.\ref{j32_c3}, it is clear that trigonal distortion has two remarkable effects. First, compared with the undistorted phase diagram in Fig.\ref{pd_undistortion}, it is evident that the distortion drives the system across the Mott transition (indicated by the dashed line) at rather smaller critical values of interaction $U_c(\lambda)$. This finding suggests that the distortion may help stabilize the TMI phase in a physically realistic range of interactions before the system undergoes a transition to a magnetically ordered phase at strong interaction. The considerable decrease of the critical Mott transition point can be traced back to the effect of distortion on the $j=3/2$ manifold. Without distortion this manifold represents four degenerate states which in turn contribute to the formation of bands. However, upon the inclusion of distortion this degenerate manifold splits into two Kramers pairs separated by amount of energy related to the strength of the distortion, {\em i.e.} $\Delta_3$. The corresponding bands will also be separated by the same energy scale. Thus with distortion, we are dealing with a half-filled band, with an effective bandwidth is reduction. So, a smaller Hubbard interaction is needed for the Mott transition.\cite{kim:prl08} 

Second, distortion stabilizes the TMI phase by extending its region of the phase diagram in comparison with the small region seen around $U\approx 9$ in the undistorted lattice. (See Fig.\ref{pd_undistortion}.)  We note that the GMI phase is found at both small and large spin-orbit coupling in the presence of a trigonal distortion.

The lower panel in Fig.\ref{j32_c3} explicitly shows the effect of distortion at fixed interaction $U=4t$, which is relevant to the application of pressure, for example. At small distortions $\Delta_3 \lesssim t$, most of the phase diagram is dominated by metallic and STI phases.  One can think of distortion as a driving parameter that transfers system from the rotor condensed phase (below the dashed line) into uncondensed phase (above the dashed line). Although the actual form of the geometrical distortion could be more complicated than the one we considered here, the result is appealing as this minimal distortion can drive the system across a variety of phases. Starting from the STI phase at zero distortion, the ground state of the system can exhibit a metallic behavior or perhaps transits to GMI and TMI phases with increased distortion. We hope this observation will help motivate new classes of experiments searching for exotic quantum phases in correlated materials with strong spin-orbit coupling.

\subsection{Compression and Elongation of the Oxygen Octahedra}
In this subsection we study the effects of the second type of distortion, Eq.\eqref{c4}, which describes a tetragonal distortion of the octahedron along one of its axes. This distortion preserves the ${C_4}$ rotation of an octahedron about the elongated axis, say the $z$-axis in Fig.\ref{pyrochlore}b. At zero spin-orbit coupling, the degeneracy of the $t_{2g}$ manifold will be lifted by this distortion. Compression of the octahedron, $\Delta_4 >0$, lowers the energy of the  $d_{xy}$ orbital (which is at zero energy by our convention) below that of the doubly degenerate $d_{yz}$ and $d_{zx}$ orbitals, with energy $\Delta_4$. An elongation (expansion), $\Delta_4<0$, of an octahedron lowers the energy of  the $d_{yz}$ and $d_{zx}$ orbitals relative to $d_{xy}$. This rearranging orbitals strongly affects the magnetic properties of the double perovskites in the strong interaction limit.\cite{dodds:arxiv10} 

When spin-orbit coupling is present, the levels split in a more complicated way.  Similar to Eq.\eqref{c3_zeeman} for the $e'_g$ manifold, the spin-orbit coupling results in the following effective Hamiltonian for a proper linear combination of $|yz\rangle$ and $|zx\rangle$ states:\cite{jin:arxiv09}
 \bea
  \label{c4_zeeman}\langle H_{so}\rangle_{yz,zx}=-\frac{\lambda}{2}\tau^{z}\otimes\sigma^{z}, 
  \eea 
  where $\tau$ acts within the doublet $\{-\frac{1}{\sqrt{2}}(|yz\rangle+i|zx\rangle),\frac{1}{\sqrt{2}}(|yz\rangle-i|zx\rangle)\}$, and $\sigma$ is the usual Pauli matrix of real spin. Note that the spin-orbit coupling acts like a Zeeman coupling so that the effective magnetic field has opposite direction in different states of the doublet, therefore, the time-reversal symmetry is preserved. [Evident as well from the Hamiltonian \eqref{hubbard}.]

The effect of distortion on the spin-orbit basis can also be treated in the same way. In particular, we consider its effect on the quadruplet $j=3/2$ manifold. (See Appendix \ref{app:SOC}.) With distortion the following states are obtained:
\bea \label{jz-32}
|\psi_{1} \rangle&=&-\frac{1}{\sqrt{2}}(|yz\uparrow\rangle+i|zx\uparrow\rangle),\nonumber \\
|\psi_{2} \rangle&=&\frac{1}{\sqrt{2}}(|yz\downarrow\rangle-i|zx\downarrow\rangle),
\eea
with energy $\varepsilon_{1,2}=\frac{1}{2}(2\Delta_4-\lambda)$ and
\bea \label{jz-12}
|\psi_{3} \rangle=C\left[-f(\lambda,\Delta_4)|yz\downarrow\rangle-if(\lambda,\Delta_4)|zx\downarrow\rangle+|xy\uparrow\rangle\right],\nonumber \\
|\psi_{4} \rangle=C\left[f(\lambda,\Delta_4)|yz\uparrow\rangle-if(\lambda,\Delta_4)|zx\uparrow\rangle+|xy\downarrow\rangle\right],\nonumber \\
\eea
with energy $\varepsilon_{3,4}=\frac{1}{4}(2\Delta_4+\lambda-\sqrt{4\Delta_4^{2}+4\Delta_4\lambda+9\lambda^{2}})$. Here $C$ is a normalization constant depending on $f$. Note that in Eq.\eqref{jz-12} $f$ is a function of its arguments with $f\to1$ as $\Delta_4\to0$. In the limit of vanishing distortion, the above sates $|\psi_{1,2}\rangle$ and $|\psi_{3,4}\rangle$ reduce to the four sates $|\frac{3}{2},\pm \frac{3}{2}\rangle$ and $|\frac{3}{2},\pm \frac{1}{2}\rangle$ of the quadruplet $j=3/2$ manifold, respectively. Note the $|\psi_{1,2}\rangle$ keeps the character of $j^{z}=\pm \frac{3}{2}$ states even for $\Delta_4 \neq 0$. With electron occupation $n_d=3$ and for $\Delta_4>0$ ($\Delta_4<0$), the states $|\psi_{1,2}\rangle$ ($|\psi_{3,4}\rangle$) form a half filled band, and we will see that the $C_4$ distortions strongly affect the phase diagram found in Fig.\ref{pd_undistortion} for the undistorted lattice.

\begin{figure}
\begin{center}
\includegraphics[width=7.7cm]{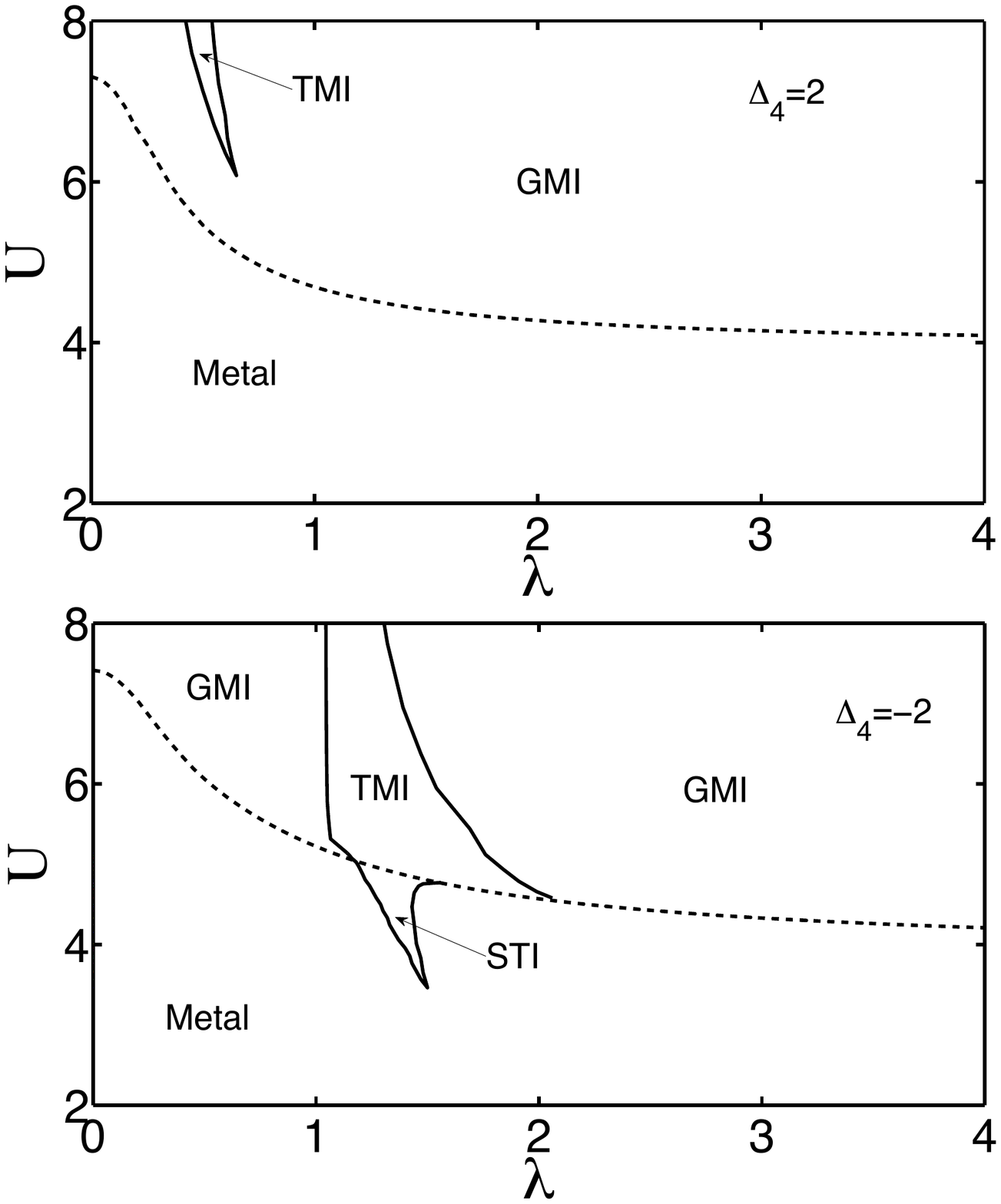}
\caption{Phase diagram of the $j=3/2$-band model with $C_4$ tetragonal distortion. The upper and lower panel correspond to compression and elongation distortion of octahedra, respectively. We set $\Delta_4=2t$ for compression and $\Delta_4=-2t$ for elongation.  All energies are expressed in units of $t$.  In both phase diagrams the dashed line separates the rotor uncondensed phase (above) from the condensed phase (below).  Note that for these values of distortion, the topological phases are interaction-driven, {\em i.e.} they do not extend down to the non-interacting limit as they do in Figs. 2-4.} \label{j32_c4}
\end{center}
\end{figure}

Fig.\ref{j32_c4} depicts the phase diagram of the $j=3/2$ model with compression (elongation) distortion in upper (lower) panel. While both compression and elongation of octahedra possess almost the same critical (dashed) line for the Mott transition, the topological phases occupy rather different regions. For example, the case of compression does not support a STI phase, while elongation does.  For strong compressional distortion $\Delta_4>0$, the bands are mainly comprised of the states $|\psi_{1,2}\rangle$, which are not spin-orbital {\em entangled}. The states $|\psi_{3,4}\rangle$, on the other hand, are spin-orbital entangled states. Even for elongation, most partions of the condensed phase (below the dashed line) is dominated by the metallic phase even in the presence of interaction. 

For both signs of the $C_4$ distortions, strong interactions open up a gap in the bulk spinon spectrum turning the GMI phase into the TMI phase as shown Fig.\ref{j32_c4}.  For $\Delta_4>0$, a  weaker distortion may extend the boundary of the TMI phase towards weaker interactions, and perhaps open a region with a STI phase. This is because at weaker distortion there would be a considerable contribution from unentangled states in the formation of the bands. This is clearly seen in the lower panel of Fig.\ref{j32_c4}, where a finite region with a STI phase is established. If the spin-orbit coupling is kept fixed, at very strong distortion the coefficient $f$ in Eq.\eqref{jz-12} tends to zero, and therefore the states become unentangled. This may partly explain why at small interaction the metallic phase is dominant. However, the lower panel of Fig.\ref{j32_c4}  reveals that interactions can drive the formation of strong topological insulators even when the STI phase is not present in the non-interacting limit, as it is in most of the cases considered previously. A two dimensional analogue of this problem has been studied elsewhere,\cite{Raghu:prl08,jun:prb10,liu:prb10,Sun:prl09,Zhang:prb09,Varney:prb10} where it is shown that the interaction-driven insulating phases with nontrivial topology can be found on a variety of different lattices.

While the distortion favors the metallic phase in the weak interaction limit and the GMI phase in the strong limit (above the Mott transition) compared to the non-distorted case (Fig.\ref{pd_undistortion}), it is possible that strong disorder can transform the metallic phase and GMI into topological phases, the so called topological Anderson insulator.\cite{Jain:prl09,Groth:prl09,Guo:prl10} In particular, the effect of disorder on the GMI phase would be an interesting problem.

\section{Summary and Conclusions}
\label{conclusions}

In this work we investigated the phase diagram of some transition metal oxides with $5d$ orbitals on the pyrochlore lattice.  We focused on the interplay between electron correlation, spin-orbit coupling, and distortion.  Our main results are summarized in the phase diagrams presented in Figs. \ref{j12_ph}-\ref{j32_c4} obtained within the slave-rotor mean-field theory.  

Examples of pyrochlore transition metal oxides include $\mathrm{A_2Ir_2O_7}$ (A=Y, Pr, Eu or other rear-earth elements), $\mathrm{Cd_2Os_2O_7}$, and $\mathrm{Cd_2Re_2O_7}$, in which different transition ions favors either $j=1/2$ or $j=3/2$ manifolds to be partially occupied.\cite{mandrus:prb01,singh:prb02}  A central feature of our work was to considered distortion of local oxygen octahedra surrounded the transition ion in the presence of interactions. Such distortions are inherent to the systems we studied.\cite{yang:prb10,subramanian:chem83}

We first studied the effect of trigonal distortion on the $j=1/2$ phase diagram already obtained in Ref.[\onlinecite{Pesin:np10}]. In the noninteracting limit, we found the distortion destabilized the STI phase and turned it into a metallic phase.\cite{yang:prb10} (See Fig.\ref{j12_ph}.) However, a $\Delta_3<0$ can also help  stabilize a weak topological insulating (WTI) phase which becomes a weak topological Mott insulator (WTMI) above the Mott transition line.  To the best of our knowledge, these features have not been obtained in previous interacting models before.

We also extended the study of interacting topological insulators in transition metal oxides to include the case where the $j=3/2$ manifold is partially filled. (See Fig.\ref{pd_undistortion}.) We found that strong spin-orbit coupling opens a gap in the non-interacting spectrum and the STI appears, along with a metallic phase at small spin-orbit coupling. These phases persist in the presence rather large interactions interaction due to the large band width of the non-interacting model, and eventually at large enough interactions the Mott phases appear. Most portions of the Mott phase are identified as a gapless Mott insulator (GMI). However, at some intermediate regime of spin-orbit coupling, $1\lesssim \lambda \lesssim2$, the TMI phase is obtained. 

Trigonal distortion extends the TMI phase to a wider range of interaction and spin-orbit coupling. (Fig.\ref{j32_c3}.) Trigonal distortion also decreases the critical interaction for the Mott transition. Moreover, we showed that the distortion can serve as a tuning parameter in which the transition between a variety of phases could occur by distorting the lattice, though a more realistic form of distortion could have a more complicated evolution of the phases. We also examined the effect of tetragonal distortion of octahedra caused by an elongation or compression of an octahedra along one of its axis. (Fig.\ref{j32_c4}.) The STI phase is found to be very delicate with respect to this type of distortion, and most of the phase diagram is occupied by either the metallic phase the the GMI. For strong enough distortion, however, interaction can restore both STI and TMI phases.  The restoration of these phases is an example of ``interaction-induced" topological phases as these phases do not persist down to zero interaction.

One might wonder to what extent the slave-rotor mean-field results should be trusted.  Is there an alternative method that can be used to obtain a TMI phase, for example?  As we mentioned earlier, quantum Monte Carlo methods applied to the 2-d Hubbard model on the honeycomb lattice seem to suggest\cite{Hohenadler:arxiv10,Zheng:arxiv10,Yamaji:arxiv10} that there is an intermediate, gapped phase that lives over a similar region of the phase diagram that the slave-rotor method predicts a 2-d TMI (recall that the slave-rotor method is not expected to be reliable in 2-d).  This may suggest that there is indeed  a state with fully gapped bulk excitations, but with gapless spin excitations on the boundary.  However, it may be that the slave-rotor method fails to correctly capture the collective nature of the ``true" low-energy spin-excitations by forcing them into a single-particle mean-field formalism.  One may also ask about the reliability of the slave-rotor method more generally.  The original work of Florens and Georges\cite{florens:prb02,florens:prb04} on Hubbard models shows a favorable comparison with dynamical mean-field theory and Gutzwiller projection for quantities like the quasi-particle weight and effective mass below and just above the Mott transition.  Finally, the references contained in the work of Pesin and Balents\cite{Pesin:np10} provide further support for the reliability of the slave-rotor method when compared with path-integral renormalization group calculations and variational cluster methods on frustrated lattices.  Taken together, it seems the method works reasonably well in situations where interactions are not too strong and no magnetic order is expected.  Nevertheless, a more careful study of the possibility of a TMI phase within a more sophisticated class of calculations remains highly  desirable and we hope this work will help to inspire such studies. 

Regarding the physics of pyrochlore oxides, a number of interesting directions for future study remain.  For example, it would be highly desirable to have a better understanding of the specific form of lattice distortions that occur in nature and what their influence is in terms of candidate topological phases.  It would also be interesting to obtain a better understanding of disorder on the interplay of correlations, spin-orbit coupling, and lattice distortions.  Finally, we note that even more exotic possibilities exist for novel phases when certain conditions are met.\cite{Maciejko:prl10,Swingle:arxiv10,Cho:arxiv10}  An improved understanding of how likely the conditions for these ``fractional" phases with non-trivial ground state degeneracy are to be met in real materials would be welcome.

As we restricted ourselves to interactions that were not too strong,  we did not invoke the possible magnetic phases that could be more favorable at very strong Coulomb interaction. The magnetic phase is interesting in its own right as the pyrochlore lattice has a geometrically frustrated structure. The latter property along with the proximity to the metallic phases can enhance the quantum fluctuations. Hence, even the nearest-neighbor antiferromagnetic interaction may stabilize a spin-liquid phase on the pyrochlore lattice.\cite{canals:prl98,canals:prb00,Bergman_2:prb06}  Besides the antiferromagnetic interaction, $j=1/2$ magnetic models that include some additional interactions such as Dyzaloshinsky-Moriya and other anisotropic interactions can help sustain ordering on the pyrochlore lattice at low temperatures.\cite{elhajal:prb05,champion:prb03} However, the situation is more complicated for the $j=3/2$ model: because of orbitally-dependent exchange, biquadratic (forth order in spin operators) and triquadratic (sixth order in spin operators) interactions arise.\cite{chen:arxiv10}  These new interactions give rise to some exotic phases in double perovskites,\cite{chen:arxiv10} and tetragonal distortion of octahedra can result in an anisotropic pseudo-spin antiferromagnetic exchange Heisenberg model.\cite{dodds:arxiv10} Such models can be developed for our model with distortion, too. Indeed, what magnetic phases become favorable and how they are related to the topological phases we addressed here are interesting open problems.

\acknowledgements
We thank Andreas R\"uegg and Dmytro Pesin for enlightening discussions. We gratefully acknowledge financial support from ARO Grant W911NF-09-1-0527 and NSF Grant DMR-0955778.

\appendix
\section{Transformation to orbital angular momentum states}
\label{app:L_z}
In terms of the $t_{2g}$ states, the effective $l=1$ angular momentum states are given by
\begin{eqnarray}
|l_z=1\rangle&=&-\frac{1}{\sqrt{2}}(|yz\rangle +i |zx\rangle),\nonumber \\
|l_z=0\rangle&=& |xy\rangle,\nonumber\\
|l_z=-1\rangle &=&  \frac{1}{\sqrt{2}}(|yz\rangle -i |zx\rangle),\nonumber
\end{eqnarray}
so that in terms of the $t_{2g}$ orbitals the effective $l_z$ angular momentum is given by
\[
l_z=\frac{1}{\sqrt{2}}\left(\begin{array}{ccc}
0 & 0 & 0\\
0 & 0 & i\\
0 & -i & 0\\ \end{array}\right),
\]
in the basis ($d_{xy},d_{yz},d_{zx}$).  This immediately gives $l_z^2=n_{yz}+n_{zx}$ in terms of the $t_{2g}$ state occupations.

\section{Spin-orbit coupled states}
\label{app:SOC}
The transformation between the spin $s=1/2$ in the effective $l=1$, $t_{2g}$ orbital basis ($d_{xy},d_{yz},d_{zx}$) and the basis of $|j,j_z\rangle$ for $j=1/2,3/2$ is given by
 \[
\left(\begin{array}{c} |{1\over 2},{1\over 2}\rangle \\ |{1\over 2},-{1\over 2}\rangle\\|{3\over 2},{3\over 2}\rangle\\|{3\over 2},{1\over 2}\rangle\\|{3\over 2},-{1\over 2}\rangle\\|{3\over 2},-{3\over 2}\rangle\\ \end{array} \right)=\left(\begin{array}{cccccc}
0 & \frac{1}{\sqrt 3} &  0 & \frac{i}{\sqrt 3} & \frac{1}{\sqrt 3} & 0 \\
\frac{1}{\sqrt 3} & 0 &  \frac{-i}{\sqrt 3} & 0 & 0 & \frac{-1}{\sqrt 3}\\
 \frac{1}{\sqrt 2} &  0 & \frac{-i}{\sqrt 2} & 0 & 0 & 0\\
0 & \frac{-1}{\sqrt 6} &  0 & \frac{-i}{\sqrt 6} & \frac{2}{\sqrt 6} & 0\\
\frac{1}{\sqrt 6} & 0 &  \frac{-i}{\sqrt 6} & 0 & 0 & \frac{2}{\sqrt 6}\\
0 & \frac{1}{\sqrt 2} &  0 & \frac{-i}{\sqrt 2}  & 0 & 0\\
 \end{array}
\right)
\left(\begin{array}{c} |yz \uparrow\rangle \\ |yz \downarrow\rangle\\|zx \uparrow\rangle\\|zx \downarrow\rangle\\| xy \uparrow \rangle\\|xy \downarrow \rangle \end{array} \right).
\]


%

\end{document}